\documentclass{ws-ijmpa}

\def\draftnote{
\small September 23, 2005\quad PITHA~05/16\quad\\ 
\hspace*{0.23cm} 
\mbox{Talk presented at the International Conference on QCD and Hadronic
Physics, Beijing, China, June 16--20, 2005}}

\begin{document}

\markboth{M. Beneke}{Hadronic B decays}

\catchline{}{}{}{}{}

\title{HADRONIC B DECAYS WITH QCD FACTORIZATION}

\author{M. Beneke}
\address{Institut f\"ur Theoretische Physik E, RWTH Aachen\\
D--52056 Aachen, Germany}

\maketitle


\begin{abstract}
A brief summary of outstanding theoretical issues and recent 
results from the QCD factorization approach to exclusive hadronic 
$B$ decays is provided.
\end{abstract}

\keywords{B decays; factorization}

\section{INTRODUCTION}

Hadronic two-body decays of $B$ mesons play a central role in the 
on-going programme to clarify the sources of CP violation and rare 
flavour-changing processes. Where experimental information is not 
sufficient to determine strong interaction amplitudes from the data 
themselves, the matrix elements $\langle f|O_i|\bar B\rangle$ (with $O_i$ 
an operator from the effective weak interaction and $f$ a two-body 
final state) must be provided by theory.

QCD factorization\cite{Beneke:1999br} is a synthesis of the heavy
quark expansion ($m_b\gg\Lambda_{\rm QCD}$) with soft-collinear 
factorization for hard processes (particle energies $\gg\Lambda_{\rm QCD}$) 
to compute the matrix elements 
$\langle M_1 M_2|O_i|\bar B\rangle$ in an expansion in $1/m_b$ and 
$\alpha_s$. Only the leading term in $1/m_b$ assumes a simple 
form. The basic formula is 
\begin{eqnarray}
\label{fact}
\langle M_1 M_2|O_i|\bar B\rangle 
&=& F^{B M_1}(0)\int_0^1 \!du\,T_i^I(u)
\Phi_{M_2}(u) \nonumber\\[0.0cm]
&&\hspace*{0cm}
+ \int \!d\omega du dv \,T_i^{II}(\omega,u,v)\,\Phi_B(\omega)\Phi_{M_1}(v) 
\Phi_{M_2}(u),
\end{eqnarray}
where $F^{BM_1}(0)$ is a standard heavy-to-light form factor at
$q^2=0$, $\Phi_{M_i}$ and $\Phi_B$ are light-cone distribution 
amplitudes, and $T_i^{I,II}$ are perturbatively calculable 
hard-scattering kernels. ($M_1$ and $M_2$ are light mesons, and $M_1$ is 
the meson that picks up the spectator antiquark from the $\bar B$ meson.) 
The formula shows that there is no long-distance interaction between the
constituents of the meson $M_2$ and the $(B M_1)$ system at leading 
order in $1/m_b$. This is the precise meaning of factorization. 

\section{THEORETICAL ISSUES}

\subsection{Factorization of spectator scattering}

The factorization argument for the first term on the right-hand
side of (\ref{fact}) is very similar to the case when $M_1$ 
is a heavy meson.\cite{Beneke:1999br,Bauer:2001cu} The second 
term is more interesting, since it involves a hard interaction 
with the soft spectator antiquark in the $\bar B$ meson, which 
introduces the scale $(m_b\Lambda_{\rm QCD})^{1/2}$. The detailed 
structure of this term has now been clarified with the help of 
soft-collinear effective theory (SCET).\cite{Bauer:2000yr,BCDF}

It is a consequence of ``colour transparency'' that the meson 
$M_2$ (which does not pick up the spectator quark) factorizes from 
the $(B M_1)$ system already when the scale $m_b$ is integrated 
out. Thus one obtains (matching to SCET${}_{\rm I}$)
\begin{eqnarray}
\label{fact2}
\langle M_1 M_2|O_i|\bar B\rangle 
&=& \int_0^1 \!\!du \,\Phi_{M_2}(u) \,\bigg\{T_i^I(u) \,F^{B M_1}(0)
+\!\int \!d\tau \,C_i^{II}(\tau;u) \,\Xi^{B M_1}(\tau;0)\bigg\}, 
\quad
\end{eqnarray}
where $\Xi^{B M_1}(\tau;0)$ denotes a new bi-local 
form factor related to the SCET${}_{\rm I}$ matrix element 
$\langle M_1|\bar\xi(0) \!\not\!\!A_{\perp c}(s n_+) h_v(0)| 
\bar B\rangle$. The same function appears in the 
factorization formula for the heavy-to-light form 
factor.\cite{Bauer:2004tj} We can therefore apply the 
factorization argument for form factors\cite{Beneke:2003pa} to 
write $\Xi^{B M_1}(\tau;0)$ as (matching to SCET${}_{\rm II}$) 
\begin{eqnarray}
\label{J}
\Xi^{B M_1}(\tau;0) = \int d\omega dv\,J(\tau;\omega,v) 
\,\Phi_B(\omega)\Phi_{M_1}(v),
\end{eqnarray}
which implies (\ref{fact}) with $T_i^{II}(\omega,u,v) 
=\int d\tau \,C_i^{II}(\tau;u)\,J(\tau;\omega,v)$. The 
short-distance function $J(\tau;\omega,v)$ contains the scattering 
with the spectator antiquark at the scale 
$(m_b\Lambda_{\rm QCD})^{1/2}$. Using (\ref{J}) provides 
more predictive power, since $\Xi^{B M_1}(\tau;0)$ is otherwise 
unknown, but is not mandatory, since $M_2$ factorizes 
at the scale $m_b$. This is exploited in what is sometimes 
(inappropriately) called the SCET approach to hadronic $B$ 
decays.\cite{Bauer:2004tj} 

Has factorization for $B$ decays to two light mesons been proven?
SCET provides precise prescriptions for the extraction of the various 
short-distance functions and up to now all results are in agreement 
with (\ref{fact}). The remaining issues are more fundamental. SCET 
factorization ``proofs'' often tend to assume what 
should be part of the proof, namely that SCET {\em is} the correct 
framework. This requires an investigation of the analytic structure 
of Feynman integrals in QCD. The observation that  
soft-collinear factorization in SCET${}_{\rm II}$ is in general 
not maintained in loop corrections due to the absence of a regulator 
that respects factorization\cite{Beneke:2003pa,Becher:2003qh} 
provides an example of the possible pitfalls in SCET factorization 
``proofs''. 

\subsection{Radiative corrections}

The short-distance functions $T^I_i$, $T^{II}_i$ are currently 
computed up to order $\alpha_s$. This means that the strong-interaction 
phases (hence, direct CP asymmetries) and spectator-scattering effects 
are known only at leading order (LO). A next-to-leading order (NLO) 
calculation of these quantities is particularly 
important in view of phenomenological evidence (mainly from the 
$\pi\pi$ and $\pi K$ final states) that the strong-interaction phases 
and the ratio of the colour-suppressed to colour-allowed tree 
amplitude, $C/T$ (or $\alpha_2/\alpha_1$), might be larger than 
the (LO) factorization result. 

A complete NLO calculation of phases and spectator-scattering 
is not yet available. The NLO correction 
to the function $J(\tau;\omega,v)$ has been 
computed,\cite{Hill:2004if,Beneke:2005gs,Kirilin:2005xz} and 
has been found to increase $C/T$.\cite{Beneke:2005gs} A partial 
calculation of the NLO penguin contributions to $T^{II}_i$ has 
been performed.\cite{Li:2005wx} The completion of these calculations 
will answer the question whether the apparent discrepancies  
between LO factorization and data have a short-distance 
explanation. 

\subsection{Power corrections}

\subsubsection{Scalar penguins}

The only $1/m_b$ power correction for which there is unambiguous 
evidence is the (pseudo)scalar penguin contribution to the QCD penguin 
amplitude ($a_6$). The scalar penguins arise from the Fierz 
transformation of the V+A penguin operators in the effective weak 
Hamiltonian and their interference with the V-A penguin contribution 
strongly discriminates between final states of two pseudoscalar
mesons (PP), one pseudoscalar and one vector meson (PV or VP), 
and two vector mesons (VV). Fortunately, the leading 
scalar penguin contribution is calculable despite being a
power correction, and its inclusion into the factorization 
formula (\ref{fact}) is mandatory for a successful 
phenomenology.\cite{Beneke:1999br,Muta:2000ti,Cheng:2000hv,Du:2001hr,Beneke:2002jn,Aleksan:2003qi,Beneke:2003zv,deGroot:2003ms}
An understanding of the factorization properties of this power
correction would provide a better justification for this procedure,
but has not yet been completed. A preliminary investigation\cite{BN} 
indicates that factorization is not preserved, but the
factorization-breaking terms appear to be numerically suppressed. 
An enumeration of power-suppressed diagram topologies related 
to this question can also be found in Ref.~\cite{Feldmann:2004mg}.

\subsubsection{Weak annihilation}

Weak annihilation constitutes a class of $1/m_b$ power corrections
which does not factorize as in (\ref{fact}). A phenomenological 
model in terms of a single parameter, which also
allows for a long-distance strong-interaction phase, is often 
used\cite{Beneke:1999br} to account for this contribution. In this model 
the single most important annihilation amplitude comes from  
penguin operators, and it is indistinguishable from the QCD penguin
amplitude.  There is evidence from data that an additional 
contribution to the QCD penguin amplitude is required that 
could be ascribed to weak annihilation, and which is compatible 
with the assumed size of this contribution (up to 25\% of the penguin
amplitude for final states of two
pseudoscalar mesons and significantly larger for vector meson final
states) in the standard parameterization.\cite{Beneke:2003zv} 
An estimate of the annihilation amplitude with QCD sum 
rules\cite{Khodjamirian:2005wn} leads to numbers in a similar 
range.  In the absence of higher-order calculations of the QCD penguin
amplitude the data does not allow to draw the conclusion that the weak
annihilation mechanism is significant at all. Evidence for this 
comes from the annihilation-dominated decay $\bar B_d\to D_s^+ K^-$ 
with an observed branching fraction in agreement with the estimate 
from the phenomenological model.\cite{Battaglia:2003in}
A better quantitative control of weak annihilation is nevertheless 
one of the key issues in the theory of hadronic $B$ decays, in
particular for final states containing one or two vector mesons.

\subsubsection{3-particle contributions}

Some power corrections related to $q\bar q g$ light-cone distribution 
amplitudes have been evaluated.\cite{Yang:2003sg,Yeh:2005gy}

\subsubsection{Charm penguins} 

Charm penguins, that is, contractions that contain a charm quark 
loop, have been discussed for some time as a source of potentially 
large corrections.\cite{Ciuchini:2001gv} While these discussions 
centered on the issue whether charm penguins could induce numerically 
large power corrections, it has recently been
claimed\cite{Bauer:2004tj} that charm penguins may not factorize  
at all even at leading power in contradiction to (\ref{fact}). 
To be precise, the short-distance part of the charm penguin loops  
is part of the standard calculation of the penguin amplitudes. The 
issue is whether charm quark loops have (incalculable) long-distance 
contributions that do not vanish as $m_b$ goes to infinity. The 
argument says that charm loops have a significant contribution 
from the charm threshold region, in which non-relativistic 
power counting must be applied and factorization is not guaranteed. 

There appears to be a confusion about how the non-relativistic 
treatment of charm is combined with the heavy quark expansion 
in $m_b$.\cite{Beneke:2004bn} If $m_b\to \infty$ at fixed $m_c$ 
(for instance $m_c v^2\sim \Lambda_{\rm QCD}$), the charm 
quark is like a light quark compared to the scale $m_b$ 
and there is no principal difference between light and charm 
quark loops, which both factorize. If $m_b\to \infty$ 
at fixed $m_c$ then also the smallest non-relativistic scale 
$m_c v^2\gg \Lambda_{\rm QCD}$, the threshold region becomes 
perturbative, and standard perturbative non-relativistic resummation 
applies.\cite{Beneke:1999zr} The non-perturbative remainder is related to 
a power correction parameterized by the matrix element of a 
higher-dimensional operator. It follows that the charm penguins 
factorize since the long-distance contribution always carries a factor 
$1/m_b$.\cite{Beneke:1999br}

\section{\mbox{EVIDENCE FOR (AND PROBLEMS WITH) FACTORIZATION}}

\subsection{Evidence}

The good overall agreement of the calculated branching fractions 
with observations, in particular with the parameter set S4 defined 
in Ref.\cite{Beneke:2003zv} provides clear evidence that the leading 
amplitudes (colour-allowed tree and QCD penguin) are approximately 
correctly obtained with factorization. Perhaps the most 
important evidence of factorization comes from the observed 
non-universality of the QCD penguin amplitude between pseudoscalar 
and vector meson final states. In factorization there is a strong 
correlation with the $J^{PC}$ quantum numbers of the primary 
weak current, which leads to $M_1 M_2$ QCD penguin amplitudes 
roughly as follows:
\begin{eqnarray}
\mbox{PP} \sim \underbrace{a_4}_{\mbox{V}\mp\mbox{A}}+
\underbrace{r_\chi a_6}_{\mbox{S+P}},
\qquad 
\mbox{PV} \sim a_4 \approx  \frac{\mbox{PP}}{3},
\qquad
\mbox{VP} \sim a_4-r_\chi a_6 \sim -\mbox{PV}.
\end{eqnarray}
The suppression of the PV and VP penguin amplitudes relative to PP 
is responsible for the smaller PV branching fractions, and the 
interference of penguin amplitudes explains the surprisingly 
different branching fractions of 
$B\to\eta^{(\prime)}K^{(*)}$.\cite{Beneke:2002jn,Lipkin:1990us} 
The relevance of factorization 
is further corroborated by the non-observation of large direct 
CP asymmetries, since it is a feature of the heavy-quark limit 
that the strong interaction phases are suppressed. 

\subsection{Problems}

A more detailed comparison also reveals difficulties with
factorization. The infamous wrong-sign 
prediction\cite{Beneke:1999br} of the direct CP asymmetry 
in $B\to\pi^\mp K^\pm $ 
is an example. This discrepancy is actually quantitatively a small 
effect, which may disappear due to  higher-order 
radiative corrections or a strong phase in the weak annihilation 
amplitude.  There also exist indications of a large colour-suppressed 
tree amplitude. It remains to be seen whether this can 
be explained by spectator-scattering 
alone.\cite{Beneke:2005gs,Beneke:2003zv} A general observation 
is that the prediction of strong phases can be rather 
uncertain not only because there is currently no NLO prediction, 
but also because a power correction $\Lambda_{\rm QCD}/m_b$ 
is parametrically not much smaller than $\alpha_s(m_b)$. 
Nevertheless, it is difficult to see how factorization could explain 
strong phases around $90^\circ$ of a QCD penguin or colour-suppressed 
tree amplitude (relative to the colour-allowed tree) as are sometimes 
reported. Obviously, before giving up a wonderful theory one 
would like to see the experimental data improve to the point 
that the case for very large phases can be made with certainty.

\section{THREE (EXEMPLARY) ROUTES TO CKM PARAMETERS}

\subsection{$(\bar\rho,\bar\eta)$ from $B\to\pi\pi, \pi K$}

A global fit of $(\bar\rho,\bar\eta)$ to the $B\to\pi\pi, \pi K$ branching
fractions in QCD factorization has already been performed in 
Ref.\cite{Beneke:1999br} and later updates gave results in good
agreement with the standard unitarity triangle fit. The most recent
fit,\cite{Charles:2004jd} which includes CP asymmetries, 
gives $\gamma =(62^{+6}_{-9})^\circ$ to be compared to 
$\gamma =(62^{+10}_{-12})^\circ$ from the standard fit. It is prudent
to regard this result with caution, since it arises from averaging
measurements that individually give rather different values for 
$\gamma$.

\subsection{$\gamma$ from time-dependent CP asymmetries in 
$b\to d$ transitions}

Given a calculation of the penguin-to-tree amplitude ratio, $\gamma$
(or $\alpha$) can be determined directly from the sin-oscillation of 
the time-dependent CP asymmetry in $B\to\pi^+\pi^-$, $S_{\pi\pi}$. 
The determination is rather accurate, since the strong phases $\delta$ 
enter only at second order in $\delta$. The analogous 
quantity $S_{\pi\rho}=(S_{\pi^+\rho^-}+S_{\pi^-\rho^+})/2$ in 
$B\to \pi^\pm\rho^\mp$ decays is particularly clean, since the 
penguin amplitude is significantly smaller (see above). The 
experimental values $S_{\pi\pi} = 0.13\pm 0.13$ and 
$S_{\pi\rho}=-0.50\pm 0.12$ provide clear evidence for the penguin
contribution, since otherwise $S_{\pi\pi}=
S_{\pi\rho}=-\sin 2(\beta+\gamma)$. Given $\beta$ and the 
calculation of the penguin amplitude, one finds\cite{Beneke:2003zv} 
$\gamma = (70^{\,+8}_{\,-8})^\circ$ (from $S_{\pi\rho}$) and 
$\gamma = (66^{\,+13}_{\,-12})^\circ$ (from $S_{\pi\pi}$) 
in nice mutual agreement and with the standard fit. 

\subsection{CKM parameters from $B\to\rho\gamma$}

QCD factorization also applies to exclusive radiative 
decays.\cite{Beneke:2001at,Bosch:2001gv}
Recent analyses of branching fractions, CP- and isospin asymmetries 
in $B\to\rho\gamma$ decays have shown that $(\bar\rho,\bar\eta)$ 
can be extracted from $b\to d\gamma$ transitions 
alone.\cite{Ali:2004hn,Bosch:2004nd,Beneke:2004dp} Although 
not competitive in accuracy with the standard fit, this provides 
another consistency check for the CKM mechanism.  
In particular $\mbox{Br}(B^0\to\rho^0\gamma)<0.4\cdot 10^{-6}$ 
implies $\left|V_{td}/V_{ts}\right|<0.21$, which already 
cuts into the range allowed by the standard fit. 

\section{PUZZLES, NOW AND THEN}

\subsection{The $B\to\pi K$ puzzles}

Certain unexpected features of ratios of $B\to\pi K$ branching 
fractions and CP asymmetries have become preeminent in 
2003\cite{Beneke:2003zv,Yoshikawa:2003hb,Gronau:2003kj,Buras:2003yc} 
and have since then been extensively discussed. The most
straightforward interpretation seems to be an enhanced  
electroweak $b\to s$ penguin amplitude with a large CP-violating
phase, and an enhancement of the colour-suppressed tree amplitude. 
See the talk of R. Fleischer at this conference for a detailed 
discussion. 

\subsection{Sin$\,(2\beta)$ from $b\to s$ transitions}

The time-dependent CP asymmetry in $b\to s$ penguin transitions 
is expected to be close to $\pm \sin(2\beta)$, since 
$b\to c\bar c s$ and $b\to s\bar s s$ 
have (nearly) the same weak phase, and subleading amplitudes 
are CKM-suppressed. 
Calculations of $\Delta S_f \equiv S_{f(b\to s)} - 
\sin(2\beta)_{J/\psi K_S}$ in QCD factorization 
confirm this expectation and yield a positive 
$\Delta S_f$ for $f=\pi^0 K_S,\eta^\prime K_S,\phi K_S,\omega K_S$, 
which is very small ($\approx 0.02$) for 
$f=\eta^\prime K_S,\phi K_S$.\cite{Beneke:2003zv,Cheng:2005bg,Beneke:2005pu} 
This is in contrast to data which upon averaging over final states 
give $\Delta S_f =-0.19\pm 0.07$. 

\subsection{$B\to\eta^{\prime}K$}

The decay $B\to\eta^{\prime}K$ has the largest branching fraction
among all charmless $B$ decays. Moreover, there is an 
interesting pattern in related final states: 
$\mbox{Br}(\eta^\prime K) \approx 20 \,
\mbox{Br}(\eta K) \mbox{ but } \mbox{Br}(\eta^\prime  K^*)<
\mbox{Br}(\eta  K^*)$. The QCD factorization analysis of 
these decays\cite{Beneke:2002jn} reveals several new decay 
mechanisms for final state mesons with flavour-singlet components, 
and explains this pattern as an interference of QCD 
$b\to s$ penguin amplitudes, which are different for PP and PV 
final states and can have different signs for $\eta$ and 
$\eta^\prime$ owing to their different strange content. 
In particular, $\mbox{Br}(B\to\eta^\prime K)\approx 70\cdot 10^{-6}$ 
is in the central range of theoretical results, which however carry 
a large uncertainty. 

\subsection{Polarization in $B\to VV$}

Decays to two vector mesons offer additional dynamical information 
due to the possibility of polarization studies. 
In the heavy-quark limit both vector mesons are longitudinally 
polarized, and one expects the hierarchy $A_0 \gg A_- \,\gg A_+ $ 
of helicity amplitudes. Transverse polarization is a power correction.
This expectation is not always borne out by the data, since for 
instance the longitudinal polarization fraction is only 
about 0.5 for $B\to\phi K^*$. No anomaly is observed for
tree-dominated decays. The resolution of this ``polarization puzzle''
in the context of QCD factorization is due to the 
observation\cite{Kagan:2004uw} that weak annihilation makes a very
large contribution to the VV penguin amplitude. A large depolarization
in penguin-dominated decays is therefore not in contradiction with 
theoretical estimations. Unfortunately, the theoretical predictions 
are also very uncertain due to the lack of a sensible theory for 
annihilation effects. 

\section*{Acknowledgements}

This work is supported in part by the 
DFG Sonder\-forschungs\-bereich/Trans\-regio~9 
``Com\-puter\-ge\-st\"utz\-te Theoretische Teilchenphysik''.

\end{document}